# FRACTAL COLD GAS AS DARK MATTER IN GALAXIES AND CLUSTERS

Daniel Pfenniger
Observatoire de Genève, CH-1290 Sauverny, Switzerland

Françoise Combes
DEMIRM, Observatoire de Paris
61 Av. de l'Observatoire, F-75 014 Paris, France

**Abstract.** The conjecture that dark matter in galaxies is mostly cold fractal gas exposed in another paper at this conference is developed in the more general context of the thermodynamics of the ideal isothermal gas subject to gravitational instability. This simple gas model already contains the contrary ingredients able to prevent an asymptotic equilibrium: any growing gravothermal singularity evaporates in a finite time, and any tendency to uniform gas is gravitationally unstable. The paradox is simply resolved by allowing fractal states, which are then scale-free and steady *in average*, but non-differentiable and time-dependent.

If we apply to clusters the lessons learned with galaxies, we are led to the conclusion that gas in clusters at a temperature much below the virial temperature should also adopt a fractal structure and become inhomogeneous. The same instrumental biases acting at galactic scale and preventing the detection of the smallest and coldest sub-resolution clumps in the fractal are then even more relevant for cluster gas measurements.

The large baryonic mass observed in the cluster hot gas and the morphology-density relation suggest also gaseous dark matter in spirals. If this dark gas component remains undetected in spirals, the same instrumental biases should hold in clusters, where cooling hot gas disappears from detection near the centre.

# 1. EVOLUTION OF GALAXIES AND GALACTIC DARK MATTER

Nowadays many hints point toward a secular evolution of galaxies much faster than classically envisioned in the 60's[1]. In particular at the scale of galaxy clusters or galaxy groups the morphology-density relation (e.g. [2]) and the Butcher-Oemler effect indicate possible major modifications of galaxies in much less than the Hubble time. At the scale of galaxies the better understanding of mergers[3], galaxy dynamics[4], and the observations of metal rich and blue bulges[5] also indicate that galaxies are still changing today, with probable sub-Gyr phases of evolution. If the Hubble sequence is an evolutionary sequence, then the only reasonable evolutionary sense is from Sm-Sd to Sa, because all the irreversible processes such as bulge formation, star formation and heavy elements enrichment are increasing this way[6].

Now the $M/L$ ratio along the Hubble sequence decreases from more than 100 in dwarf irregulars to nearly normal for visible matter ($\approx 10$) in Sa's. Admitting regular processes dark matter cannot be removed selectively during galaxy evolution because the total mass of galaxies increases from Sd to Sa (e.g. [7]). Also it is dynamically ruled out to accrete more than a few percents of matter within an already formed stellar disk without heating or destroying it[8],[9]. But it remains possible to accrete large amounts of angular momentum rich and dissipative matter at the disk periphery, as in high resolution cosmological simulations with gas (e.g. [10]). So the simplest hypothesis is to assume that during the secular evolution of isolated disk galaxies dark matter transforms into stars. The only way to achieve this is to have most dark matter in some gaseous form beyond the optical disks.

The small amount of dark matter in Sa galaxies and stellar disks puts a constraint on the amount of non-gaseous dark matter, such as brown dwarfs, black-holes, or other dark matter candidates in the original mix making late-type spirals. Since the amount of dark matter in early-type spirals is at most of the order of the visible matter, it means that in late-type spirals the fraction of non-gaseous dark matter is less than about 10%.

The obvious problem with gaseous dark matter is then to explain why the mass of gas as usually determined from HI and CO observations is, although proportional to dark matter, insufficient by a large factor ($\approx 20$). By critically reviewing the knowledges we have about interstellar gas, we can state the following[11]: interstellar gas is typically very inhomogeneous and multi-phase, covering decades of density and temperature ranges; in gravitating systems such as galaxies only gas at a temperature higher than the virial temperature ($\approx 10^4$ K in galaxies, $\approx 10^7$ K in clusters) tends to be smooth. Colder gas tends to be highly lumpy, without necessarily forming immediately stars or Jupiters (since it is still there!). This is indicative that gravitation rules the clumpiness of the gas by the Jeans' instability condition. Also observations of the local ISM show that most of the commonly measured gas mass is in a cold form and fills only a few percents of the volume, an unusual fluid indeed.

With today's observational constraints it appears that a large range of a priori possible clumps at subparsec scales can hardly be detected. These clumps might be even colder, down to 2.7 K, than those we can presently measure ($\gtrsim 5$ K). Neglecting the highly clumpy state of cold gas at sub-resolution scale is evaluated to account for the systematic discrepancy between dark matter and HI in the outer gaseous galactic disks[11].

# 2. THERMODYNAMICS OF THE JEANS UNSTABLE IDEAL GAS

Several major problems of astrophysics would appear clearer if the following idealised problem would be first understood: *What is the asymptotic state of an ideal gas (made of small perfectly elastic particles) contained in a box and maintained at a temperature sufficiently low to reach the gravitational (Jeans) instability?* Before introducing more complex physical ingredients, it is important to understand this simple model, because otherwise one might miss important aspects of cosmic gas physics.

In a spherical container of radius $r$ the equilibrium state of a hot enough gas is the isothermal sphere (see e.g. [12], p. 500). However, below a critical temperature ($kT < 0.4\,GMm/r$), no equilibrium state does exist. ($M$ is the total gas mass, and $m$ the particle mass.) Some parts may temporarily collapse and heat indefinitely (the gravothermal catastrophe), but this departure from thermodynamical equilibrium diverges toward a state equivalent to a finite mass isolated system, which evaporates completely in a finite time; this occurs by increasing the central density and temperature, but also by decreasing the total collapsing mass (see e.g. [12], p. 523). This contrary behaviours between the beginning and end of a singularity show that in a such Jeans unstable isothermal gas any tendency to collapse is soon prevented by evaporation, but also any tendency to uniformity is prevented by Jeans' instability. The asymptotic state can only be dynamical.

On the other hand, we expect also that the average behaviour should become scale-free, since no scale between the largest and smallest sizes (given by the box and the particle) is involved in the process. Of course in space no wall confines gas, yet the problem remains because an isothermal sphere requires then an infinite mass. Having a finite mass the whole gas blob evaporates in about 100 crossing times (see e.g. [12], p. 525). So the paradox persists, in no way a cold ideal gas in isothermal conditions can find an asymptotic state of rest.

A solution to this paradox is suggested by observations and simulations. Molecular clouds are observed to be fractal over at least 4 decades, from 100 down to 0.1 pc[13],[14], and shearing sheet experiments of self-gravitating and Jeans unstable layers develop long-range and time-dependent correlations[15] which appear typical of fractals. The asymptotic state is indeed scale-free but dynamical: a fractal state in which clumps exist at every scale, but collide, merge, or disrupt so that the statistical ensemble is *in average* steady and scale-free. The thermodynamics of fractal systems is today in the infancy (e.g. [16]).

A fractal state for interstellar gas is very different from the classical picture of fiducial round clouds of determined scale. Not only it has never been possible to show that such clouds are stable over a long time, but also the particular retained size of these clouds has been chosen for purely convenience reasons. In the literature several attempts have been made to "deconstruct" the fiducial interstellar cloud model, yet despite its obvious crudeness, the model is still alive by lack of alternative picture.

## 3. REAL COLD COSMIC GAS

In practice no gas is ideal at every scale, and there always exists an upper and lower scale at which the geometrical fractal model breaks. A largest scale in a galaxy is determined by differential rotation ($\sim (d\ln\Omega/dr)^{-1}$), but the smallest scale must be determined by the microphysics. Without invoking complications such as magnetic fields, a simple possibility has been given[11] involving first order opacity properties of ideal gas in a black-body radiation field[17]: the scale at which there is a transition from the isothermal to the adiabatic regime. In cold conditions near 3 K, for a mix of $H_2$ and He this transition occurs at solar system sizes, with a corresponding mass of the order of Jupiter. This mass is proportional to $T^{1/4}\mu^{-9/4}$, so is slowly variable with $T$, but more strongly dependent on the molecular weight $\mu$. Interestingly, above the $H_2$ and H ionization temperatures $\mu$ decreases from about 2.3 to 0.63, and the critical mass increases by a factor $\sim 30$, from about 0.003 $M_\odot$ at $T \approx 2000$ K to about 0.1 $M_\odot$ at $T \approx 8000$ K[11], crossing rapidly the brown dwarf regime, which suggests a link with the stellar IMF cutoff below 0.1 $M_\odot$.

Recently, Ferland et al.[18] have computed in detail the cooling of uniform gas in cluster conditions taking into account many coupling channels of matter with radiation. They find that indeed cooling may proceed efficiently from millions of K down to 3 K.

In star formation problems it is usually assumed that an adiabatic proto-star cloud collapses in isolation. Yet we have estimated that on the contrary, if the cold gas has a fractal dimension $D \approx 1.5 - 2.0$, as derived from the empirical Larson relations $M \sim r^D$ in Galactic molecular clouds, the collision time is similar or shorter than the crossing time at every scale, so the isolation hypothesis does not hold[11]. The clump collisions are then slightly supersonic, which is appropriate to disrupt clumps trying to collapse further in the adiabatic regime. At higher temperature, above the H-ionization, the fractal dimension $D$ becomes smaller than 1, and then the collision time increases allowing deeper adiabatic collapses (assuming initially the empirical properties of the fractal gas in the Galaxy).

## 4. CLUSTER EVOLUTION AND DARK MATTER

The above considerations have some applications to clusters. Taking the observational facts at face: the morphology-density relation shows that nearly all spiral galaxies are transformed into early-type galaxies or completely destroyed when crossing clusters. The high hot gas mass fraction found in clusters implies that dark matter in spirals is also at least for such a fraction in the form of gas, presumably cold. If cold gas in spiral galaxies is able to escape detection, the same can be expected to happen in clusters. The most natural scale ($\approx 10 - 50 \, \mathrm{AU}$) of cold gas at low temperature is hardly accessible with present instruments. Since cluster dark matter profiles follow hot gas profiles except near the center, it is therefore likely that a substantial fraction of the cluster dark matter is also in the form of cold fractal gas. The higher fraction of dark matter in the dense conditions at cluster centers would result naturally from the higher cooling rate of the hot gas becoming cold 3 K gas.